\documentclass[american,aps,pra,reprint,superscriptaddress, longbibliography]{revtex4-1}
\usepackage[T1]{fontenc}
\usepackage[latin9]{inputenc}
\usepackage{color}
\usepackage{babel}
\usepackage{amsthm}
\usepackage{amsmath}
\usepackage{amssymb}
\usepackage{graphicx}

\usepackage[unicode=true,pdfusetitle, bookmarks=true,bookmarksnumbered=false,bookmarksopen=false,  breaklinks=false,pdfborder={0 0 0},backref=false,colorlinks=false] {hyperref}
\hypersetup{ colorlinks,linkcolor=myurlcolor,citecolor=myurlcolor,urlcolor=myurlcolor}

\makeatletter
\@ifundefined{textcolor}{}
{%
 \definecolor{BLACK}{gray}{0}
 \definecolor{WHITE}{gray}{1}
 \definecolor{RED}{rgb}{1,0,0}
 \definecolor{GREEN}{rgb}{0,1,0}
 \definecolor{BLUE}{rgb}{0,0,1}
 \definecolor{CYAN}{cmyk}{1,0,0,0}
 \definecolor{MAGENTA}{cmyk}{0,1,0,0}
 \definecolor{YELLOW}{cmyk}{0,0,1,0}
 }
\theoremstyle{plain}

\theoremstyle{plain}

\ifx\proof\undefined
\newenvironment{proof}[1][\protect\proofname]{\par
\normalfont\topsep6\p@\@plus6\p@\relax
\trivlist
\itemindent\parindent
\item[\hskip\labelsep
\scshape
#1]\ignorespaces
}{%
\endtrivlist\@endpefalse
}
\providecommand{\proofname}{Proof}
\fi
\theoremstyle{plain}

\providecommand{\lemmaname}{Lemma}
\providecommand{\definitionname}{Definition}
\providecommand{\propositionname}{Proposition}

\usepackage{babel}
\usepackage{txfonts}
\usepackage{colortbl}\definecolor{myurlcolor}{rgb}{0,0,0.7}
\newcommand{\tr}{{\operatorname{Tr\,}}}

\def\ket#1{| #1 \rangle}

\def\braket#1{\langle  #1 \rangle}

\newcommand{\ketbra}[2]{|#1\rangle \langle#2|}

\newcommand{\haH}

\newtheorem{theorem}{Theorem}

\newtheorem{definition}[theorem]{Definition}

\definecolor{orange}{RGB}{255,127,0}

\begin{document}
\title{Quantum Thermodynamics Allows Quantum Measurement Almost Without Collapse}
\author{Mohit Lal Bera}
\affiliation{ICFO -- Institut de Ci\`encies Fot\`oniques, The Barcelona Institute of Science and Technology, ES-08860 Castelldefels, Spain}
\author{Manabendra Nath Bera}
\email{mnbera@gmail.com}
\affiliation{Department of Physical Sciences, Indian Institute of Science Education and Research (IISER), Mohali, Punjab 140306, India}

\begin{abstract}
We introduce a quantum measurement process that is capable of characterizing an unknown state of a system almost without disturbing or collapsing it. The underlying idea is to extract information of a system from the thermodynamic quantities like work(s) and heat in a process, thereby uncovering a fundamental correspondence between information and thermodynamics. We establish an improved notion of \emph{information isolation} and show that a process is isolated if it respects the \emph{first law} of quantum thermodynamics for a given set of conserved quantities or charges. The measurement process involves a global unitary evolution of the system, an apparatus, and a battery which supplies work(s). The global unitary respects the first law. The full information about the system is accessed by counting the charge-wise work costs to implement the reduced evolution on the system and the apparatus. After the work costs are determined, the process is undone where a state of the system is retrieved arbitrarily close to the original initial state. The measurement process is also capable of characterizing an unknown quantum operation. Fundamentally, our findings make an important step towards resolving the paradoxes arising from the quantum measurement problem, such as the Wigner's friend paradox, and the issue related to the objective reality of quantum states. We discuss the technological implications of the almost collapse-free measurement process. 
\end{abstract}

\maketitle

\section{Introduction}
Collapse or disturbance induced in a quantum system during a measurement process is considered to be one of the least understood features and the fundamental trait of quantum mechanics \cite{Wheeler16, braginsky92, Busch16}. One may recall the example of Schr\"odinger cat \cite{Schrodinger35}, where the cat is in a superposition between dead and alive states. By the act of observation in the measurement process, the cat is collapsed into either dead or alive state. Similarly, by observing the particle nature of an electron, the wave nature is destroyed in a double-slit experiment \cite{Davisson27}. The very act of observation inevitably collapses the cat state or the electron state with a certain probability. Furthermore, the observer can selectively induce the collapse at her will in the measurement process. Thus, the information extracted in the measurement only relates to this collapsed or modified state, and it is observer-dependent. This aspect raises the inter-related questions: (i) whether there is an objective reality of the quantum state, i.e., if it can be measured independent of an observer, and also, (ii) if quantum mechanics is intrinsically probabilistic or random, etc. Since the inception of quantum mechanics, these questions have been explored in great detail \cite{Wheeler16, Busch16, ZurekRMP03, SchlosshauerRMP05}. Except for some recent insights, e.g. \cite{PBR12, Colbeck12, Horodecki2015, Budiyono17}, the questions related to objective reality, measurement induced collapse, and probabilistic nature of quantum mechanics are not resolved yet.

There have been substantial efforts put forward to minimize the disturbance induced in a system during a measurement process. These result in various measurement protocols, e.g., quantum non-demolition measurement \cite{Braginsky80, Braginsky96}, weak measurement \cite{Aharonov88, Dressel14}, etc. However, it is now widely believed that one cannot learn about a state of a quantum system without disturbing or collapsing it, and more information one wants to acquire about a system more disturbance she has to introduce in it.

\begin{figure}
\includegraphics[width=0.85\columnwidth]{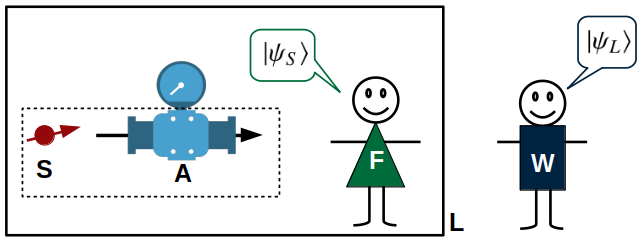}
\caption{The thought experiment leading to Wigner's friend paradox \cite{Wigner67}. An experimenter $F$ (Wigner's friend) performs a quantum measurement using an apparatus $A$ to determine the state of a quantum system $S$. After an interaction between system and apparatus, $F$ infers about the system by observing only the apparatus $A$. With that, she inevitably introduces an irreversible modification, i.e., collapse, in the quantum system $S$. Therefore the system $S$, by definition, is transformed via a non-unitary process. All these take place inside an ``isolated'' laboratory $L$. In principle, all the systems inside $L$ (i.e., $S$, $A$, and $F$) are quantum mechanical in nature. Being outside $L$, an observer $W$ (Wigner) finds that the entire laboratory $L$ is evolving in ``isolation'', i.e., following a unitary process. He has no ``information'' at all on whether $F$ has performed a measurement or not. Now $W$ performs a measurement on $L$ to characterize the quantum state of the system $S$, and his inference on the state of $S$ appears something different from what his friend $F$ makes. Therefore, two observers $F$ and $W$, being inside and outside the isolated laboratory $L$, have different inferences on the same system $S$. This leads to a contradiction, as well as the paradox.
\label{fig:WignerParadox}}
\end{figure}

To examine how objective and consistent is the assumption of measurement induced collapse, Wigner proposed a thought experiment in 1967. This led to the famous ``Wigner's friend paradox'', and there he argued that ``quantum mechanics cannot have unlimited validity'' \cite{Wigner67}. In this thought experiment  (see Figure \ref{fig:WignerParadox}), Wigner's friend $F$ performs a measurement to learn about a quantum system $S$ inside an ``isolated'' laboratory $L$. Thereby, she introduces a collapse or disturbance in the system $S$ which is not a unitary process. While being outside, Wigner ($W$) finds that the entire laboratory $L$ evolves unitarily. He performs a measurement on the entire $L$ to characterize the state of the system $S$. However, $W$'s inference appears different from $F$. This leads to the paradox and it is exclusively due to the measurement induced collapse in $S$. Recently, this paradox has been extended further to claim that ``Quantum theory cannot consistently describe the use of itself'' \cite{Frauchiger18}.

Here, we make an attempt to address the quantum measurement problem, and to do that we reconsider Wigner's friend paradox. By critically analyzing the measurement process and the assumptions therein, we demonstrate that the assumptions taken by Wigner \cite{Wigner67} and the authors in \cite{Frauchiger18} are incomplete. We invoke the quantum thermodynamical notion of information isolation, based on a deeper relation between information and thermodynamics, and exploit these to study thermodynamical aspects of the measurement process, which was missing so far. In particular, we relate the flow of information with the changes in thermodynamic quantities, such as different types of work. Using this insight we devise an almost reversible measurement protocol with which a quantum system is characterized by the thermodynamic quantities associated with the measurement process. The process is almost without collapse, or almost collapse-free, in the sense that it introduces an arbitrarily vanishing disturbance in the system being measured. This leads to an important step toward resolving the measurement paradoxes, as well as the quantum measurement problem. 

\section{Isolation vs information isolation}
One important assumption Wigner made in his thought experiment \cite{Wigner67}, as well as the authors in \cite{Frauchiger18}, is that the laboratory $L$ is isolated from the rest of the universe as long as the $L$ undergoes a unitary process. What does this isolation mean? Does it imply that there is no exchange of information with the outside universe? No, not necessarily! To understand the notion of isolation, we invoke (quantum) thermodynamics here and consider that the laboratory $L$ is surrounded by an environment. If we assume that information can only be exchanged between $L$ and the environment through some ``entropic'' mode via open-quantum system dynamics, then Wigner is very much correct. In fact, the entropic mode of information exchange happens through nothing but \emph{heat} exchange. It requires $L$ to have exclusive access to the environment in order to exchange quantum system(s) or exclusive access to some degree of freedom of the environment to establish correlation so that the system locally undergoes a non-unitary evolution. As long as the $L$ is closed, i.e., it does not have exclusive access to the environment, such a heat exchange cannot take place. Interestingly the entropic mode is not the only means, and information can flow via non-entropic modes too. In fact, $L$ can exchange \emph{work}, e.g. energy, with the environment which can in principle carry information! We call it work, as the energy exchange takes place without an entropy flow or heat exchange. Indeed, the system needs to have access and interact with the environment in order to undergo a unitary evolution and, at the same time, exchange energy. In that case, there is a possibility that the system and the environment build up a correlation among themselves. Then, locally, the environment modifies its entropy. But, this happens without an entropy flow, and it is only due to the fact that the correlation between the system and environment is ignored. The system's local entropy remains unchanged as it still evolves unitarily. As long as the correlation is taken into account, there is no heat flow from the system to the environment as such \cite{Bera16}, although the environment locally modifies its entropy. Therefore, a unitarily evolving system only exchanges work.

\begin{figure}
\includegraphics[width=0.65\columnwidth]{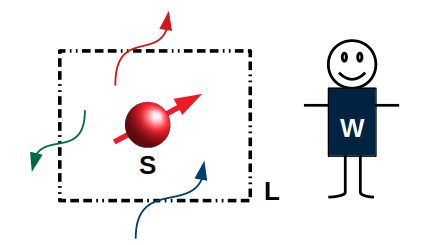}
\caption{The notion of information isolation. A system $S$ is enclosed in a laboratory $L$. No heat exchange is possible with the environment, as the system can only undergo unitary evolution. However, such evolution can lead to the exchange of charges with its environment, as indicated by the arrows. These charges could the three spin angular momentums corresponding to the Pauli spin matrices for a qubit system. By counting this flow of charges, an outside observer $W$ can, in principle, infer about the evolution. Thus, as far as information exchange is concerned, $L$ is not isolated from the outside environment, even though the $S$ is not allowed to ``interact'' with the environment. The system is in isolation only when it undergoes unitary evolution and, at the same time, there is no exchange of charges with the environment.
\label{fig:Isolation}}
\end{figure}

Let us consider a two-level system $S$, with the Hamiltonian $H_S=E_0 \ketbra{0}{0} + (E_0 +\Delta) \ketbra{1}{1}$, which is initially in the ground state $\ket{0}$. The system $S$ is enclosed in $L$, as shown in Figure \ref{fig:Isolation}, and an observer $W$ does not have ``access'' to it. But, $W$ has full access to the rest of the universe, i.e., environment. Assume that $W$ knows the initial state of the system $S$ and has a scheme to quantify the energy change of the environment without affecting $S$. Now say, $W$ notices a decrease in energy of amount $\Delta$ in the environment. Since $L$ is closed and only undergoes unitary evolution, the energy transferred to the system is in pure form, which can be considered as the work cost of the unitary evolution inside $L$ paid by the environment. With this, $W$ can deterministically conclude that $S$ has undergone a transition $\ket{0}\rightarrow \ket{1}$. Here he assumes that the total energy of the universe is conserved, as guaranteed by the \emph{first law} of thermodynamics. Thus, the exchange of energy or work can indeed carry information.

However, for some process, the information may not be accessed via the exchange of energy. Say, the $S$ goes through a unitary evolution $\frac{1}{\sqrt{2}}(\ket{0} + \ket{1}) \rightarrow \frac{1}{\sqrt{2}}(\ket{0} - \ket{1})$. The observer $W$ would not be able to guess if any process has taken place inside $L$, as this does not lead to an energy change in the environment. But $W$ may choose a different observable corresponding to a conserved quantity, say $H_S^\prime=\ketbra{0}{1} + \ketbra{1}{0}$, and apply \emph{first law}  corresponding to that observable. Here we term the quantity of interest, that is conserved in the processes, as the \emph{charge}. The transformation will alter the charge value of the system, and $W$ can, in principle, be able to access some information about the process from the change of the same charge in the environment. Albeit, the charge observable may or may not commute with the system Hamiltonian.

Nevertheless, what we argue here is that the outside observer $W$ can in principle learn about the process happening inside $L$ by counting the charge flow. Consequently, \emph{Wigner's assumption of isolation, saying $W$ cannot have any information about the process inside a closed laboratory $L$, is incomplete!} In fact, as we shall show later, there always exists a set of charges and a protocol with which $W$ can almost perfectly characterize the process happening inside $L$ and also the quantum state of $L$. As a consequence, $W$ will not only get to know if $F$ has performed measurement inside $L$ or not but also the almost complete information about the process. Hence, there will be no contradiction between the inferences made by $F$ and $W$ in Wigner's thought experiment (see Fig. \ref{fig:WignerParadox}). This argument is instrumental to resolve the Wigner's friend paradox \cite{Wigner67}.

Suppose, the outside observer $W$ has access to the charges $\{Q_\alpha\}_{\alpha=1}^c$ only, where $c$ represents the number of charges. When can we say that the process in $L$ is occurring in isolation from $W$? In other words, how can we guarantee that $W$ cannot have access to any information about the process via either the entropic mode or non-entropic modes given the charges? The answer is given in the following definition.

\begin{definition}[Information isolation]\label{thm:Isolation}
A system, with the charges $\{Q_\alpha\}_{\alpha=1}^c$ and the corresponding charge observables $\{\widehat{Q}_\alpha\}_{\alpha=1}^c$, is undergoing a process. Then it is in isolation from an outside observer with respect to the charges if, and only if, the process is driven by a unitary $U$ and it satisfies the first law of quantum thermodynamics, i.e.,
\begin{align}\label{eq:IsolationCommu}
 [U, \ \widehat{Q}_\alpha]=0, \ \forall \ \alpha=1,\ldots,c,
\end{align}
where each charge is strictly conserved by the process.
\end{definition}

As we have discussed earlier, the unitarity of the process guarantees that there is no information exchange via entropic mode, i.e., heat. This is an adiabatic process in the thermodynamic sense. In addition, the commutation relations \eqref{eq:IsolationCommu} imply that the unitary $U$ strictly conserves the charges $\{ Q_\alpha \}_{\alpha=1}^c$ separately. Therefore, there is no information exchange via non-entropic mode, i.e., no flow of work corresponding to the charges $\{ Q_\alpha \}_{\alpha=1}^c$ between the system and the outside universe. The reverse statement is also true. If an observer, being a part of the outside universe, cannot have access to the information about the process, then there are no exchanges of information either via the entropic mode (heat) or the non-entropic mode (work corresponding to the given charges). Clearly, the former can only be ensured if the process is unitary, i.e., adiabatic. The latter can be guaranteed if the unitary evolution strictly conserves all the charges separately, which is expressed by the commutation relations \eqref{eq:IsolationCommu}.

It is obvious that to learn about an unknown process, the outside observer $W$ has to have access to a set of different quantum observables or charges that, in general, do not commute with each other, i.e., $[\widehat{Q}_\alpha, \ \widehat{Q}_\delta ]\neq0$ for some $ \alpha, \ \delta$. As a consequence, there may not exist a closed evolution or unitary process that \emph{strictly} conserves all the charges separately. In other words, the unitary operator $U$ does not commute with all the charges simultaneously. The non-commuting charges also cannot have sharp values for an arbitrary quantum state, as dictated by the quantum uncertainty relation.

However, the first law with non-commuting charges can only be ensured in certain situations. For instance, consider an ensemble that is composed of an asymptotically large number of independent and identical systems. This is known as the \emph{asymptotic limit} or \emph{asymptotic regime}. In this regime, the non-commuting charges commute with each other on average \cite{Ogata13}. There exists a global unitary that simultaneously commutes with all the charges on average \cite{Halpern16}. Also, an arbitrary quantum state can have a sharp value for each charge, where the value is exactly equal to the expectation value of the charge. Therefore, the first law of quantum thermodynamics, which states that a quantum process on system and environment conserves all the charges simultaneously, can be ensured on \emph{average} in the asymptotic limit \cite{Guryanova16, Halpern16}. In that case, the information isolation can be guaranteed on average. The first law can also be ensured in the strict sense for a situation where the system is evolved in the presence of a large environment, where the environment acts as the reference frame and, at the same time, ensures strict conservation of the non-commuting charges \cite{Popescu19}. The strict information isolation can be ensured in this situation. However, it should be noted that implementing these first law respecting unitary operations are very difficult in general.

\section{Quantum Measurement Almost Without Collapse}
Using the thermodynamic notion of information isolation, we introduce a measurement protocol employing which information of a system can be extracted almost without disturbing or collapsing it. This is done by implementing an measurement evolution arbitrarily close to a well-defined measurement unitary operation and properly accounting the charge-wise work costs for that. The measurement protocol can also be extended to characterize an unknown quantum operation. \\

\begin{figure}
\includegraphics[width=0.75\columnwidth]{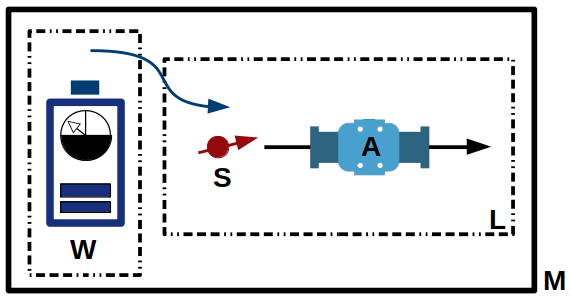}
\caption{Schematic for quantum measurement almost without collapse. A system $S$ and an apparatus $A$ are enclosed in a laboratory $L$. The battery $W$, which also acts as an observer and resides outside the $L$. In the first step of the measurement process, we choose one charge operator from the set $\{\widehat{Q}_{\alpha}^{mn} \}$ given by the observables \eqref{eq:NonLocChargeX}-\eqref{eq:NonLocChargeZ} as the Hamiltonian of $SA$. Say the chosen charge operator is $\widehat{Q}^{pq}_{\delta} \in \{\widehat{Q}_{\alpha}^{mn} \}$ corresponding to the charge $Q^{pq}_{\delta}$. The $SAW$ composite constitutes an informationally isolated system $M$ given the charge $Q^{pq}_{\delta}$ and respects total charge conservation, i.e., first law. The observer $W$ employs a well-defined unitary operation ($U_{SA}$) on $SA$ via a global unitary process applied on $SAW$. The $W$ records the work cost $\Delta \bar{q}_{W_\delta^{pq}}$ corresponding to the charge $Q^{pq}_{\delta}$ to implement this unitary. After that the transformation is reversed to retrieve the initial state of $SA$ with vanishing error. The process is repeated with for all the other charges $\{Q_{\alpha}^{mn} \}$ to count the work costs $\{\Delta \bar{q}_{W_{\alpha}^{mn}} \}$. Using these work costs, one is able to characterize the state of $S$ with a vanishing error.  See text and the Appendix for more details.
\label{fig:StateMeas}}
\end{figure}

\noindent {\it Measurement of quantum states -- } Previously we have studied the situation where we know the initial state of a quantum system and we are to learn about the process it is undergoing inside a closed laboratory (see Figure \ref{fig:Isolation}). Now we consider the reverse, where we know the process and we are to characterize the quantum state, as shown in Figure \ref{fig:StateMeas}.
\begin{table}[h]
 \begin{center}
 \begin{tabular}{ | c | c | c | }
\hline
{\bf Measurement} & {\bf Known} & {\bf Unknown} \\
\hline
{\bf Process measurement } & {Quantum state } & Quantum process \\
\hline
{\bf State measurement } & {Quantum process} & Quantum state \\
\hline  
\end{tabular}
\end{center}
\end{table}

Consider a $d$-dimensional quantum system $S$ in an unknown state, given by the density matrix,
\begin{align}\label{eq:ArbQuditState}
 \rho_S=\sum_{m,n=0}^{d-1} p_{mn} \ \ketbra{m}{n}_S.
\end{align}
The orthonormal basis set $\{\ket{m}_S \} \in \mathcal{H}_S$ in the system Hilbert space is chosen at our convenience. The goal is to acquire a complete knowledge about the state $\rho_S$ via quantum measurement. We introduce a $d$-dimensional quantum apparatus $A$ with the Hilbert space $\mathcal{H}_A$, which is initially in a pure state $\ket{0}_A$. The system and the apparatus are jointly evolved with the particular measurement unitary $U_{SA}$, given by
\begin{align}\label{eq:MeasUnitary}
 U_{SA}=\sum_{m=0}^{d-1} \ketbra{m}{m}_S \otimes V_A^{0 \rightarrow m},
\end{align}
where the $V_A^{0 \rightarrow m}=\sum_{n=0}^{d-1} \ketbra{(m+n) \ \mbox{mod} \ d }{n}_A$. The bases $\{\ket{m}_A \} \in \mathcal{H}_A$ are mutually orthonormal to each other. Note, the $U_{SA}$ represents a CNOT-gate for $d=2$. The joint initial state $\rho_{SA}^i=\rho_{S} \otimes \ketbra{0}{0}_A$ is then transformed to
\begin{align}\label{eq:MeasStateEntQudit}
 \rho^i_{SA} \longrightarrow \rho^f_{SA}=U_{SA} \ \rho_{SA}^i \ U_{SA}^\dag=\sum_{m,n=0}^{d-1} p_{mn} \ \ketbra{mm}{nn}_{SA},
\end{align}
where $\ket{mm}_{SA}=\ket{m}_{S} \otimes \ket{m}_{A} \in \mathcal{H}_S \otimes \mathcal{H}_A $. The final state $\rho_{SA}^f$ possesses maximum possible correlation, such as quantum entanglement, between the system and apparatus for the given initial state $\rho_S$ \cite{Streltsov15}. Through this correlation only, the apparatus state establishes correspondence with the system state.

In the traditional measurement process, after the (unitary) interaction, the apparatus is selectively observed or projected in various states. From the probabilities of the outcomes on the apparatus part, the information about the system state is inferred. By the act of projecting the apparatus, the system state is also projected or collapsed to a particular state. However, this happens because of the presence of quantum entanglement between the system and the apparatus in the post-interaction state. Nevertheless, there is an unavoidable disturbance introduced in the system $S$.

We depart from this traditional approach (see Figure \ref{fig:StateMeas}). We pause ourselves after the measurement unitary evolution and study it's quantum thermodynamical aspects. Since the measurement interaction is unitary ($U_{SA}$), we can consider that the state transformation is nothing but an adiabatic process where no heat is exchanged between $SA$ and its environment. However, it requires some exchange of works, corresponding to different charges, to implement the unitary $U_{SA}$. This means the environment has to perform these works on $SA$. In principle, this exchange of charges (i.e., works) carries some information about the joint initial state $\rho_{SA}^i$ of $SA$ or, equivalently, about the state $\rho_{S}$ of the system $S$.

To access information about the state $\rho_S$, we judiciously choose a set of charges that are capable of encoding all the information, including the ones residing in correlation present in the post-interaction state of $SA$. The \emph{local} charges are not enough to serve the purpose. Rather it requires charges that are \emph{non-local} in character. We propose one such set of $(d^2-1)$ charges, and the corresponding charge observables are given below.

\noindent (i) $\frac{d(d-1)}{2}$ symmetric charges, for $0 \leqslant m < n \leqslant d-1$,
\begin{align}\label{eq:NonLocChargeX}
  \widehat{Q}^{mn}_x=\ketbra{mm}{nn}_{SA} + \ketbra{nn}{mm}_{SA}.
\end{align}

\noindent (ii) $\frac{d(d-1)}{2}$ anti-symmetric charges, for $0 \leqslant m < n \leqslant d-1$,
\begin{align}\label{eq:NonLocChargeY}
 \widehat{Q}^{mn}_y=-i \ketbra{mm}{nn}_{SA} + i \ketbra{nn}{mm}_{SA}.
\end{align}

\noindent (iii) $(d-1)$ symmetric charges, for $1 \leqslant m  \leqslant d$,
\begin{align}\label{eq:NonLocChargeZ}
\widehat{Q}^{mm}_z= \sqrt{\frac{2}{m(m+1)}}\!\left( \sum_{k=0}^{m-1}\ketbra{kk}{kk}_{SA} \!-\! m\ketbra{mm}{mm}_{SA}   \!\right).
\end{align}
For notational simplicity, we denote the charge observables as $\{ \widehat{Q}^{mn}_{\alpha}\}$ with the corresponding charges $\{ Q^{mn}_{\alpha}\}$ respectively, where  $1\leqslant m=n \leqslant d$ for $\alpha=z$, and $0\leqslant m < n \leqslant d-1$ for $\alpha=x,y$. Note that we consider only $(d^2-1)$ orthonormal charges out of complete set of $(d^4-1)$  mutually orthonormal charges available for the $d^2$-dimensional Hilbert space $\mathcal{H}_S \otimes \mathcal{H}_A$.

The unitary $U_{SA}$ does not conserve each charge separately. This also implies that the unitary $U_{SA}$ leads to an informationally leaky evolution. The changes in various charges for the transformation $\rho_{SA}^f=U_{SA} \ \rho_{SA}^i \ U_{SA}^\dag$, where the $U_{SA}$ is the unitary given in Eq.~\eqref{eq:MeasUnitary}, are
 \begin{align}\label{eq:ChangeChargeSA}
 \Delta q^{mn}_{\alpha} =  \tr[ \rho_{SA}^f \widehat{Q}^{mn}_{\alpha}]- \tr[ \rho_{SA}^i \widehat{Q}^{mn}_{\alpha}].
\end{align} 
For an arbitrary state $\rho_{S}$ given in Eq.~\eqref{eq:ArbQuditState}, the charge-wise work costs to implement $U_{SA}$ are
\begin{align*}
& \Delta q_z^{mm}=\sqrt{\frac{2}{m(m+1)}}\left(\sum_{k=1}^{m-1} p_{kk}-m p_{mm}  \right), \\
&\Delta q_x^{mn}=p_{mn} + p_{nm}, \ \ \Delta q_y^{mn}= i p_{mn} - i p_{nm}, 
\end{align*}
and the quantities $\{\Delta q_\alpha^{mn} \}$ encodes complete information of the density matrix elements of $\rho_{S}$, and, thus, fully characterize the state. For instance, consider a qubit system ($d=2$) in an arbitrary state $\sigma_S=\sum_{m,n=0}^1 r_{mn} \ \ketbra{m}{n}_S$, which is to be characterized. We attach a qubit apparatus $A$ in the initial state $\ket{0}_A$. The joint initial state $\sigma_{SA}^i=\sigma_S \otimes \ketbra{0}{0}_A$ is evolved with the unitary $U_{SA}$, mentioned in Eq.~\eqref{eq:MeasUnitary}, to result in the final state $\sigma_{SA}^f$. There are three charges to characterize the state, $Q^{11}_z, \ Q^{01}_x$, and $Q^{01}_y$. The changes in charge values due to the transformation are then $\Delta q_z^{11}=-r_{11} $, $\Delta q_x^{01}= r_{01} + r_{10} $, and $ \Delta q_y^{01}= ir_{01} - i r_{10}$. Given that $r_{00}+r_{11}=1$ and $r_{01}=r_{10}^*$, one may uniquely determine the qubit state $\sigma_{S}$

The changes in the charge values in the $SA$ are compensated by the exchange of charges with the environment so that the total charges are conserved. This is the requirement imposed by the first law of quantum thermodynamics and the notion of information isolation with respect to these charges. It means that the environment performs works to implement the unitary $U_{SA}$ on $SA$. We are to determine these works which also carry the information about the state. 

Let us start with one of the charges, say $Q^{pq}_\delta \in \{Q^{mn}_\alpha \}$, and find out the cost of implementing $U_{SA}$. For that we shall consider a battery $W^{pq}_\delta$ with the charge operator $\widehat{Q}_{W^{pq}_\delta}=c^{pq}_\delta \hat{x}$, where $\hat{x}$ is the position operator and $c^{pq}_\delta$ is an appropriate constant. Now, we design a global evolution on joint space of $SA$ and $W^{pq}_\delta$ using the unitary $U_{SAW^{pq}_\delta}$ which strictly conserves total charge, i.e., $[U_{SAW^{pq}_\delta}, \widehat{Q}^{pq}_\delta + \widehat{Q}_{W^{pq}_\delta}]=0$ (see Appendix). For the initial battery state $\rho_{W^{pq}_\delta}$, the global transformation is then
\begin{align}
 \rho_{SA}^i \otimes \rho_{W^{pq}_\delta} \to \sigma_{SAW^{pq}_\delta}= U_{SAW^{pq}_\delta} (\rho_{SA}^i \otimes \rho_{W^{pq}_\delta}) U_{SAW^{pq}_\delta}^\dag.  
\end{align}
As shown in \cite{Aberg14, Malabarba15, Guryanova16}, there exists a suitable $\rho_{W^{pq}_\delta}$ for which the reduced state of $SA$ becomes, for $\epsilon >0$, 
 \begin{align}\label{eq:AlmostUnitary}
 \parallel \sigma_{SA} -  U_{SA} (\rho_{SA}^i) U_{SA}^\dag \parallel \leqslant \epsilon,
\end{align}
where $\sigma_{SA}=\tr_{W^{pq}_\delta} [\sigma_{SAW^{pq}_\delta}]$. Therefore, the $SA$ undergoes a transformation which is $\epsilon$-close to the desired (measurement) unitary transformation. Note, the $\epsilon$ can assume a value arbitrarily close to zero, thus the reduced operation on $SA$ can be made arbitrarily close to unitary $U_{SA}$. Since the $U_{SAW^{pq}_\delta}$ is strictly charge conserving, the change in charge in $SA$ is exactly opposite to the change in $W^{pq}_\delta$. Therefore, $\Delta \bar{q}^{pq}_\delta=-\Delta \bar{q}_{W^{pq}_\delta}$, where $ \Delta \bar{q}^{pq}_\delta=\tr[\sigma_{SA} \widehat{Q}^{pq}_\delta]-\tr[\rho_{SA}^i \widehat{Q}^{pq}_\delta]$ and $ \Delta \bar{q}_{W^{pq}_\delta}=\tr[\sigma_{W^{pq}_\delta} \widehat{Q}_{W^{pq}_\delta}]-\tr[\rho_{W^{pq}_\delta} \widehat{Q}_{W^{pq}_\delta}]$ with $\sigma_{W^{pq}_\delta}=\tr_{SA}[\sigma_{SAW^{pq}_\delta}]$. An important point is that once battery is detached and change in its \emph{average} charge value is recorded the initial state of the $SA$ can be recovered up to an error $\epsilon$, i.e., 
\begin{align}
\parallel \rho_{SA}^i - U_{SA}^\dag \sigma_{SA} U_{SA} \parallel \leqslant \epsilon, 
\end{align}
by separately applying the unitary $U_{SA}^\dag$ on the final state $\sigma_{SA}$ of $SA$. Thus, the reduced evolution on $SA$ induces a disturbance (or collapse) which is quantified by the value $\epsilon$. 

In addressing the fundamental inter-relation between measurement and collapse, we resort to the situation where $\epsilon \to 0$ which leads to an almost unitary evolution of $SA$, and, in turn, an almost collapse-free quantum measurement process. Then, the change in charge value in the battery becomes $\Delta \bar{q}_{W^{pq}_\delta} \to (- \Delta q^{pq}_\delta)$, where the $\Delta q^{pq}_\delta$ is defined in Eq.~\eqref{eq:ChangeChargeSA}. The same procedure is repeated with all other charges $\{Q^{mn}_\alpha \}$ and all the charge-wise work costs are determined which are arbitrarily close to the $\{- \Delta q_\alpha^{mn}\}$. Using these work costs one can characterize the unknown state $\rho_{SA}^i$ of $SA$ a vanishing error, $\epsilon \to 0$. This in turn also characterizes the unknown state $\rho_{S}$ of the system $S$. 

It is worth mentioning that while determining the \emph{average} energy change in the battery, we need to resort to the i.i.d. setting. In this setting, we need to have access to arbitrarily large number of identical copies ($N \to \infty$) of the system, apparatus, and battery composite; the measurement process is executed on each copy of the composite; perform another set of measurements only on the batteries to determine its average energy change. Therefore, the almost collapse-free measurement protocol presented above requires asymptotically large number of identical composite. \\

\noindent {\it Measurement of quantum operations:}
Analogous to the quantum states, an unknown quantum operation (or quantum channel) can also be measured almost without disturbance or collapse. Consider an arbitrary quantum operation $\mathcal{E}$ acting on a $d$-dimensional Hilbert space. The measurement protocol to fully characterize this operation involves two steps. First, (i) we prepare a bipartite maximally entangled state of the form $\ket{\psi}_{BS}=\frac{1}{\sqrt{d}} \sum_{k=0}^{d-1} \ket{kk}_{BS}$, where $\ket{kk}_{BS}=\ket{k}_{B} \otimes \ket{k}_{S} $ and $\{\ket{k}_{B}\}$ ($\{\ket{k}_{S}\}$) forms a complete set of orthonormal basis of the Hilbert space of the sub-system $B$ ($S$). Then we employ the quantum operation on the subsystem $S$ to result in the state $\rho_{BS}=\mathbb{I} \otimes \mathcal{E} (\ketbra{\psi}{\psi}_{BS})$. This state represents the Choi-Jamiolkowski matrix that has a one-to-one correspondence with the operation $\mathcal{E}$ \cite{Nielsen00}. (ii) We characterize the final state $\rho_{BS}$ following the protocols outlined above for quantum states. This, in turn, also characterizes the quantum operation $\mathcal{E}$. 

\section{Discussion and conclusions}
An arbitrary state of a quantum system can be uniquely defined by the expectation (or average) values of a complete set of charges, e.g., the charges corresponding to the generalized Gell-Mann matrices \cite{Kimura03, Bertlmann08}. These (non-commuting) charges cannot have sharp values simultaneously, as restricted by Heisenberg uncertainty relation. However, to define a state, the expectation (or average) values are enough. But, the question is how to determine these charge values. In reality, one cannot access the absolute charge values. Rather, only accessible quantities are the changes in the charge values of the system itself, or changes in some observable of a second system, e.g., environment or batteries, which depends on the initial and final states of the first one. In the prescribed measurement protocol, we have considered the latter scheme. We attach a battery with the system-apparatus composite and employ a global unitary evolution. Here the batteries act as the second systems. We then count the changes in average charge value in the batteries due to the process in order to characterize the state of the system. 

To learn about a quantum state of a system, one has to determine the changes in several non-commuting (non-local) charges corresponding to the system-apparatus composite, i.e., work costs,  when the system and apparatus evolved with a particular measurement unitary. In doing so, we have attached batteries with the system-apparatus composite and consider one charge at a time. Then, we implement a global unitary transformation which is capable of implementing a reduced operation on the system-apparatus composite arbitrary close to the well defined unitary $U_{SA}$. This is equivalent to the case $\epsilon \to 0$, in Eq.~\eqref{eq:AlmostUnitary}. The global process assumes the validity of the first law of quantum thermodynamics, that is, the total charge is conserved. The changes in charge values in the system-apparatus composite are compensated by the counter changes in the batteries which are assumed to be the part of the environment. These changes in charge values in the batteries are the work cost of the unitary evolution on the system-apparatus composite, which is paid by the battery. Each battery is responsible for accumulating changes in only one charge at a time. The value of the work is accessed from the battery after detaching it from the system-apparatus composite. Using these works, the system state is characterized. 

As a whole, our results that establish the possibility of an almost collapse-free quantum measurement process (for $\epsilon \to 0$) are expected to make deep impacts on both the fundamental understanding of quantum mechanics and quantum technologies. We summarize them below. \\

\noindent \emph{Establishing fundamental correspondence between information and thermodynamics -- } Thermodynamics and information are intimately interconnected. The role of information in thermodynamics was initially indicated by Maxwell through the famous example of Maxwell's demon \cite{Maxwell08, Maruyama09} leading to an apparent violation of the second law. By incorporating the role of (classical) information into the framework of thermodynamics, the second law was recovered by Landauer \cite{Landauer61}. He further showed that information is physical and can be converted into thermodynamic work. This inter-link has been further extended and generalized to incorporate quantum information, including entanglement, in \cite{Bera16}. There are also efforts to understand quantum thermodynamics, in particular, the second law, from the information-theoretic perspectives, e.g. \cite{Brandao13, Horodecki13, Brandao15, Binder18, Bera17}.

So far the major focus has been to study how information and energy (or charges) play roles in thermodynamics and how these, and their combinations, can be converted into work. Here, we consider the opposite and ask, \emph{whether information about a quantum state or a quantum process can be extracted from the thermodynamic quantities, such as work(s) and heat.} We give an affirmative answer, and exploit it to devise a protocol that enables an almost collapse-free quantum measurement process. Furthermore, the fundamental notion of information isolation is now founded on (quantum) thermodynamics. For a set of accessible thermodynamic quantities, like charges, a process is informationally isolated if, and only if, the process respects the first law of quantum thermodynamics. For an informationally leaky process, e.g. the unitary evolution of the system-apparatus composite in the measurement process considered in Figure \ref{fig:StateMeas}, the information can flow both via heat (entropic mode, which is mostly considered in information theories), or/and via the flow of works correspond to the charges (non-entropic modes). By properly counting these exchange of heat and works, the process and the state are characterized with vanishing error. \\

\noindent \emph{Measurement induced collapse or disturbance seems to be the consequence of ignorance -- } It is commonly believed that a measurement on a quantum system inevitably introduces disturbance into it. The disturbance here means the collapse or irreversible modification(s) in the post-measurement system. As our findings for almost collapse-free measurement justify (for the case $\epsilon \to 0$), this disturbance or loss of information is due to the ignorance or incapability of an experimenter. These may have several origins. 

First, the experimenter ignores the energetics of the measurement process involving system and apparatus, and the environment. In other words, she does not take into account the thermodynamical aspects of the process. Considering that the process is driven by a  unitary evolution, the system-apparatus composite undergoes an adiabatic transformation which leads to the exchanges of various charges, i.e., works. By disregarding the information flow via the exchange of works, she already losses some information about the process, as well as about the systems.

Second, in a traditional measurement, the experimenter only observes the apparatus to infer about the system. Thereby, she ignores the information transferred in the quantum correlation developed in the resultant system-apparatus joint state, due to the measurement unitary. When looked from the perspective of the system, this ignorance leads to the flow of information, via entropic mode or heat, from system to the apparatus. Further, if the experimenter chooses to observe the apparatus selectively in a particular state, the system state gets modified. This phenomenon is often termed as measurement induced collapse, where a selective observation of the apparatus determines the state in which the system collapses. But, this happens essentially due to the very non-local feature of quantum correlation, like entanglement, present in the post-unitary joint state. We can argue that the observer-dependent collapse of a quantum system in the process of measurement is due to the partial and selective observation made on the apparatus in the presence of non-local quantum correlation. In the proposed protocol for an almost collapse-free measurement process, we access the information stored in correlation by choosing \emph{non-local} charges. 

Therefore, the collapse or the disturbance introduced in the post-measurement state of the system is nothing but a modification required to incorporate the ignorance that is made by disregarding the quantum correlation and the information leaking due to the charge flow. Once all these are properly accounted for, as our results indicates, there is no unavoidable collapse or disturbance in quantum measurement as such, and the initial state of the system can be retrieved with vanishing error. \\

\noindent \emph{Implications of almost collapse-free measurement -- } The probabilistic outcome in a quantum measurement has been believed to be an unavoidable and fundamental feature of quantum mechanics. This characteristic feature is often associated with the assumption that a measurement inevitably induces a collapse or disturbance in the system. This is also the origin of all quantum measurement problems and related paradoxes, e.g. Wigner's friend paradox \cite{Wigner67}. The random or probabilistic nature of the measurement outcomes is also the reason to believe that there is an ``inherent randomness'' present in quantum mechanics \cite{BeraPhilo16}. Further, due to this random nature and the observer-dependence in the measurement outcomes, it is also believed that a quantum measurement cannot be understood via a dynamical process.

In contrast, we have shown that a quantum measurement can be performed, as allowed by the first law of quantum thermodynamics, to characterize a quantum state almost deterministically. Therefore, with an extrapolation, we conclude the following. (i) Quantum mechanics is not intrinsically random. The probabilistic outcomes or indeterminacy only appear in a measurement process due to ignorance as we have discussed earlier. (ii) Since the process uses unitary evolution, the quantum measurement can be understood in terms of dynamical evolution and quantum thermodynamics. (iii) The quantum measurement related paradoxes disregard the quantum thermodynamical aspects of the process, in particular, the notion of information isolation and the flow of information via charges. For example in the thought experiment leading to Wigner's friend paradox (in Figure \ref{fig:WignerParadox}), the experimenter $F$ in the laboratory $L$ ignores both the charge(s) flow as well as the correlation between the system $S$ and apparatus $A$. This leads to an apparent collapse in the system $S$. Further, it is assumed that the outside observer $W$ cannot access information about the process inside closed laboratory $L$, which is also incomplete. In principle, the experimenter $F$ can characterize the system $S$, by implementing an almost collapse-free measurement protocol that we have introduced. At the same time, the outside observer $W$ can learn about any change (driven by unitary) happening inside $L$ by counting the change in various charge values. Once the information flow by all (thermodynamical) means and the information in correlation are properly taken into account, in principle, there will not be a contradiction in the inferences made by $F$ and $W$. Hence, our almost collapse-free measurement makes an important step towards resolving Wigner's friend paradox, as well as other paradoxes related to quantum measurement problem.  

Another important point is the question of objective reality in quantum mechanics. It is questioned with the logic that one cannot characterize a quantum state without collapsing it, and the measurement outcomes are observer-dependent. This has been precisely put forward in the famous Einstein-Podolski-Rosen (EPR) article \cite{EPR35}, as: ``If, without in any way disturbing a system, we can predict with certainty (i. e., with probability equal to unity) the value of a physical quantity, then there exists an element of physical reality corresponding to this physical quantity''. Our findings indicate that a quantum state can indeed be measured with almost certainty. Therefore, the value of a physical quantity of a system can be predicted with almost certainty which is independent of the observers and thus objective. In conclusion, quantum mechanics seems to respect realism, i.e., the objective reality of quantum states and the associated physical quantities.

The EPR paradox \cite{EPR35} and subsequently the Bell's theorem \cite{Bell64, CHSH69, Brunner14} argued that the quantum mechanics cannot respect local-realism, i.e., locality and realism. The experimental observations of the violation of Bell-inequality \cite{Aspect82, Hensen15, BigBell18} prove that quantum mechanics violates either locality or realism, or both of these assumptions. Since realism seems to be respected, we, therefore, conclude that the violation of Bell inequality is exclusively due to the non-local nature of quantum mechanics or the violation of locality assumption only. \\

\noindent {\it Copying of quantum states and operations:}
Given that a quantum state is characterized almost without disturbance, an almost perfect copy of the same also can be made. This is done in two steps, first characterizing the state using the protocol of almost collapse-free measurement and then preparing it separately. We should note that this copying is very different from the quantum cloning operation considered in \cite{Dieks82, Wootters1982}. The cloning process is done in one step and involves a universal global evolution (often unitary) that is applied jointly on the state to be cloned and a blank state. However, there does not exist a universal operation that clones arbitrary state as it violates linearity of quantum mechanics, hence the no-cloning theorem \cite{Dieks82, Wootters1982}. Using our protocol, we cannot clone an unknown state as well. But, we can copy. For that we first characterize the state, and, in the next step, we prepare the state form a blank one. Note, the latter is a state-dependent process. That means, for different states, the overall copying processes are different, and this is contrary to the cloning process. The copying process does not demand a unique or universal operation here. Hence, there is no violation of linearity of quantum mechanics. It is worth mentioning that we cannot copy a state if we have access to a finite number of systems. We have to have access to an ensemble of an asymptotically large number of identical (i.i.d.) systems so that the state characterization can be done with vanishing error. Therefore, we can copy a state only if we have access to an ensemble of asymptotically large number of identical systems. Similarly, an arbitrary quantum operation can be copied just by learning the operation almost without disturbing it and then preparing it separately. \\

\noindent \emph{Possible impacts in quantum technology -- } Many quantum information-theoretic protocols rely on the so-far accepted assumptions that quantum measurement inevitably introduces collapse while characterizing the quantum states, and unknown quantum states and operations cannot be copied.

One such example is quantum key distribution (QKD), which is used for secure quantum communications between two parties. The most important aspect of QKD is that the two communicating parties are capable of detecting the presence of a third party trying to access information about the key, and this is based on the assumption that the measurement process by the third party, in general, disturbs the system. This assumption is however not stringent as we have shown that disturbance due to measurement can be made arbitrarily small. Further, an arbitrary quantum state (as well as an arbitrary quantum operation) can be almost perfectly copied. Therefore, one should re-examine the security aspects of the QKD protocol. 

In quantum information processing, there are often unavoidable noises in the quantum channels which introduce unwanted errors in the outputs. The quantum error correction protocols \cite{Nielsen00, Devitt13, Terhal15} are developed to rectify such errors. Since an arbitrary quantum channel can now be characterized using our framework, the errors or noises introduced via the channel can be known in almost full detail. Therefore, the errors can in principle be corrected with vanishing error.

Finally, what we want to highlight here is that all the technological applications that involve quantum measurement, and all the quantum information theoretic protocols that rely on the assumptions of unavoidable disturbance due to measurement, and arbitrary quantum states and operations cannot be copied are now at stake and need to be re-investigated. However, the protocols that exclusively rely on the non-local character of quantum mechanics should not be affected by our results. \\

\noindent {\bf Acknowledgements -- } The authors gratefully thank Maciej Lewenstein, Caslav Brukner, Markus Muller, Michal Horodecki, and Christian Gogolin for making valuable comments. MLB acknowledges the Spanish Ministry MINECO (National Plan 15 Grant: FISICATEAMO No. FIS2016-79508-P, SEVERO OCHOA No. SEV-2015-0522, FPI), European Social Fund, Fundacio Cellex, Generalitat de Catalunya (AGAUR Grant No. 2017 SGR 1341 and CERCA/Program), ERC AdG OSYRIS and NOQIA, EU FETPRO QUIC, and the National Science Centre, Poland-Symfonia Grant No. 2016/20/W/ST4/00314.


\appendix
\section{Quantum measurement protocol}
In the measurement protocol, outlined in the main text, we consider (explicit) batteries in conjugation with the system-apparatus composite to ensure total charge conservation. This is the requirement of the first law of quantum thermodynamics. This conservation of charges is the main underlying physical principle that is used to demonstrate that an almost collapse-free quantum measurement can be done just by reading the amount of charge exchanged with the batteries. The exchanged charges can be considered as the costs to implement the measurement interaction between system and apparatus which leads to an evolution arbitrarily close to a unitary. Here we show that there exists a protocol where the system-apparatus composite interacts with batteries via global unitary evolution and also conserve the total charges. 
 
\subsection{Explicit battery and strictly energy conserving unitary operation \label{App:Protocol1a}}
The formalism to quantify the cost of implementing an arbitrary unitary operation has been considered in \cite{Aberg14, Malabarba15} in the context of coherence catalysis and quantum clocks. Later, the protocol has been extended to study quantum thermodynamics in the presence of multiple conserved quantities \cite{Guryanova16}. We shall briefly re-iterate the protocol below. 

Let us assume that the system-apparatus composite $SA$ has the joint Hamiltonian $H_k^{SA}=\sum_{i} E^{(k)}_i \ketbra{i^{(k)}}{i^{(k)}}$. We introduce a battery $W_k$ with the Hamiltonian $H_{W_k}=\gamma_k \hat{x}$, where $\hat{x}$ is position operator and the $\gamma_k$ is a constant with the appropriate unit. The primary purpose of the battery is to compensate the change in energy of $SA$ by updating its average position so that the total energy always remains (strictly) conserved. We also define the translation operator 
\begin{align}
 \Gamma^a_{k}=\exp[-i a \hat{p}],
\end{align}
where $a$ determines the amount of translation in the position of the battery. The $\hat{p}$ is the momentum operator; the canonically conjugate operator of $\hat{x}_k$ and acts as the generator of the translation. The translation operator acts on an unnormalized position state as $\Gamma^a_{k} \ket{x}=\ket{x+ a}$. To make the framework as general as possible, we have consider three assumptions:\\

\noindent (A1) The $SA$ and $W_k$ are initially uncorrelated.

\noindent (A2) The set of allowed operations on the $SA$ and the battery composites are global unitary operations, which respects total energy conservation, i.e.,
\begin{align}
 [U_{SAW_k}, H_k+H_{W_k}]=0.
\end{align}
This is also the mathematical expression of first law in the quantum domain, as for any initial state, $\Delta E_k + \Delta E_{W_k}=0$, where $\Delta E_k$ and $\Delta E_{W_k}$ are the change in energies of the $SA$ and the battery $W_k$ respectively.

\noindent (A3) The global unitary $U_{SAW_k}$ commutes with the translation operator $\Gamma^a_{k}$, i.e., 
\begin{align}
 [U_{SAW_k}, \Gamma^a_{k}]=0.
\end{align}
This is known as the translation invariance property of the battery. It implies that the only displacement in the position of the battery is important, not the initial position.

\subsection{Implementing a unitary and counting the work cost}
Let us now elaborate on how to implement a unitary $U_{SA}$ on an arbitrary state of $SA$, given by  
\begin{align}\label{eq:SysUniApp}
 U_{SA}=\sum_{ij} u_{ij}^{(k)} \ \ketbra{i^{(k)}}{j^{(k)}},
\end{align}
where the basis states $\ket{i^{(k)}}$ are the eigenstates of the Hamiltonian $H^{SA}_k$. In general, the unitary does not conserve energy. However, we can attach the battery $W_k$ with the $SA$ and perform a global unitary
\begin{align}\label{eq:GobUniCommutCharge}
 U_{SAW_k} = \sum_{ij}u_{ij}^{(k)} \ \ketbra{i^{(k)}}{j^{(k)}} \otimes \Gamma^{a_j-a_i}_{k},
\end{align}
where $E^{k}_i - E^{k}_j=\gamma_k (a_j - a_i)$. The global unitary is designed so that it \emph{strictly} conserves the total energy, i.e.,
\begin{align}\label{eq:TotEngCons}
 [U_{SAW_k}, \ H^{SA}_k + H_{W_k}]=0.
\end{align}
The global unitary can also be expressed in the momentum representation, as
\begin{align}
U_{SAW_k}=\int dp \ V(p) \otimes \ketbra{p}{p},
\end{align}
where $V(p)=\sum_{ij} u_{ij}^{(k)} e^{-ip(a_j-a_i)}\ketbra{i^{(k)}}{j^{(k)}}$. We apply this global unitary on the joint (uncorrelated) state of $SA$ and the battery $W_k$, so that
\begin{align*}
 \rho_{SAW_k}=\rho_{SA}^i \otimes \rho_{W_k} \to \sigma_{SAW_k}= U_{SAW_k}(\rho_{SA}^i \otimes \rho_{W_k})U_{SAW_k}^\dag. 
\end{align*}
We are interested in the post-operation state of $SA$, given by
\begin{align}
 \sigma_{SA} =\tr_{W_k} \left[\sigma_{SAW_k} \right] & = \int dp \ V(p) \rho_{SA}^i V(p)^\dag \ \braket{p|\rho_{W_k}|p}, \nonumber \\
 &= \int dp \ V(p) \rho_{SA}^i V(p)^\dag \ \mu(p).
\end{align}
The $\mu(p)= \braket{p|\rho_{W_k}|p}$ are the probability distributions in the momentum space on the initial state of the battery. These are also the mixing terms and the overall evolution on the $SA$ is given by the mixture of unitary operations. The battery only undergoes a free evolution; the final state $\sigma_{W_k}=\tr_{SA} (\sigma_{SAW_k}) $ only shifts its position without affecting the shape of the momentum distribution.   

If we want to implement the unitary $U_{SA}$ on $SA$. However, for that to happen, we need to have $V(p)=V(0)=U_{SA}$ and the initial state of the battery $\rho_{W_k}$ has to have a very narrow momentum, i.e., the $\rho_{W_k}$ has to be a delta function when expressed in the momentum bases. As shown in \cite{Malabarba15, Guryanova16}, there exists a suitable $\rho_{W_k}$ so that
\begin{align}
\parallel \sigma_{SA} - U_{SA} \rho_{SA}^i U_{SA}^\dag  \parallel \leqslant \epsilon,
\end{align}
for any $\epsilon >0$. Note, we have denoted $ \rho_{SA}^f=U_{SA} \rho_{SA}^i U_{SA}^\dag$ in the main text.  Therefore, we can implement an operation on $SA$ which is arbitrarily close ($\epsilon \to 0$) to the unitary $U_{SA}$ operation. Since the global unitary $U_{SAW_k}$ strictly conserves total energy, as ensured by the commutation relation \ref{eq:TotEngCons}, we have  
\begin{align}
 \Delta E_k + \Delta E_{W_k}=0,
\end{align}
where $\Delta E_k=\tr (\sigma_{SA} H_k) - \tr(\rho_{SA}^i H_k)$ and $\Delta E_{W_k}=\tr(\sigma_{W_k} H_{W_k}) - \tr(\rho_{W_k} H_{W_k})$.

To determine the work cost of implementing the unitary $U_{SA}$ on $SA$, the battery is detached from the $SA$ and then, the change in average energy in the battery is measured. The initial and final states of the battery are not in the eigenstates of the battery Hamiltonian $H_{W_k}$ and do not have sharp energy values. However, the change in average energy can be determined given access to an asymptotically large number ($N \to \infty$) of identical batteries. This is also known as i.i.d. setting. Thus, with access to an asymptotically large number of $SA$ and batteries $W_k$ composite, we implement the identical operation on each copy of $SA$ and $W_k$ composite; detach the batteries, and then make measurements to find out the change in average energy in each battery. Once the energy change in battery is recorded, the final state $\sigma_{SA}$ of $SA$ is evolved with the unitary $U_{SA}^\dag$, so that the resultant state becomes $\epsilon$-close to the original initial state $\rho_{SA}^i$, as 
\begin{align}
\parallel U_{SA}^\dag \sigma_{SA} U_{SA} -  \rho_{SA}^i   \parallel \leqslant \epsilon.
\end{align}
The $\epsilon$ represents the amount of disturbance introduced in the original state due to the measurement process, which can be made arbitrarily close to zero ($\epsilon \to 0$).

\subsection{The measurement process}
We are now ready to implement the \emph{almost} collapse-free measurement process outlined in the main text (for $\epsilon \to 0$). Recall that to characterize a state $\rho_S$ of a $d$-dimensional system $S$, we attach a $d$-dimensional apparatus $A$ with in the joint state $\rho_{SA}^i=\rho_{S}\otimes \ketbra{0}{0}_A$ and then implement the unitary $U_{SA}$. We aim to determine the cost of such operation for the charges $\{Q_{\alpha}^{mn}\}$, given by Eqs.~\eqref{eq:NonLocChargeX}-\eqref{eq:NonLocChargeZ} in the main text. We denote these charge operators as the Hamiltonians $\{H^{SA}_k\}_{k=1}^{d^2-1}\equiv \{\widehat{Q}^{mn}_\alpha\}$. 

For the measurement process, we first attach the $SA$ with a battery. Say, for the Hamiltonian $H^{SA}_k$, we choose $W_k$ as the battery as mentioned in the earlier sub-section. Then, we find out the average work (energy) cost of implementing an operation $\epsilon$-close to the unitary operation $U_{SA}$. We repeat the step for other Hamiltonians $\{H^{SA}_k\}$, record the work costs, and characterize the unknown state $\rho_S$ with vanishing error. 
%
\end{document}